# *GIANT AND TIME-DEPENDENT MAGNETOCALORIC EFFECT IN HIGH-SPIN MOLECULAR MAGNETS*

**F. Torres and J.M. Hernández**

Dept. Física Fonamental. Universitat de Barcelona. Diagonal 647. 08028 Barcelona. Spain

**X. Bohigas**

Dept. Física i Enginyeria Nuclear. Universitat Politècnica de Catalunya. Diagonal 647. 08028 Barcelona. Spain

**J. Tejada**

Dept. Física Fonamental. Universitat de Barcelona. Diagonal 647. 08028 Barcelona. Spain

**Abstract.-** We have measured and calculated the magnetocaloric effect in macroscopic samples of oriented high-spin molecular clusters like $Mn_{12}$ and $Fe_8$ as a function of the temperature and both the intensity and the sweeping rate of the applied magnetic field. We have observed a high magnetic entropy variation around the blocking temperature of the magnetic moment of molecules and calculated the shift of the entropy variation and cooling temperature, with the sweeping rate of the magnetic field.

In recent years, there has been increasing interest in utilizing the magnetocaloric effect in refrigeration at both room temperature and cryogenic temperatures. This has been mostly induced by the pioneering work of Shull *et. al.* [1] using superparamagnetic clusters and the recent very important discoveries by Pecharsky and Gschneider[2]. The actual situation is such that several "green" devices have been constructed to cool down from room temperature using magnetic materials[3-6]. The broad spectrum of materials that may work as refrigerants covering the temperature range from 4 K to room temperature is associated to the possibility to get ferro and ferrimagnetic materials ordered in such temperature regime[7-10]. In the range of the hydrogen and helium liquefaction temperatures, superparamagnetic materials showing blocking temperatures in this interval, compete with ferro and ferrimagnetic materials. In this paper we present results obtained in a family of magnetic nanocomposites, the so called molecular clusters, from which we have deduced the occurrence of an extremely high entropy change in the Kelvin regime.

Magnetic molecular clusters like $Mn_{12}$ and $Fe_8$ [11-12] bridge the atomic and mesoscopic scales and they have the advantage with respect magnetic nanoparticles that a macroscopic sample of molecular magnets consists of a large number of chemically identical entities that are characterized by a unique set of parameters. The high value of the spin $S = 10$ of these molecules together with their small blocking temperature around few K, suggest to test these materials for very low temperature magnetic cooling. It is important to point out that experimentally it is easy to prepare a macroscopic sample of these clusters having all molecules their easy axes parallel to each other and to the applied magnetic field.

In zero magnetic field the total magnetic moment of one oriented molecular cluster is zero. In the presence of a magnetic field $H$, applied along the easy axis of the molecules, the moment of the molecules whose magnetic moments are aligned along the positive easy axis direction, $M_+$, is uncompensated with the moment, $M_-$, of the particles having the opposite orientation. These two moments verify the following differential equations:

$$dM_+/dt = -\Gamma_+ M_+ + \Gamma_- M_-$$

$$dM_-/dt = -\Gamma_- M_- + \Gamma_+ M_+$$

(1)

being $\Gamma_\pm$ the rates of the transitions between the two possible directions of the magnetic moment of individual particles, $\Gamma_\pm = \nu_\pm \exp(-U_\pm(H)V/T)$. The molecular clusters like $Mn_{12}$ and $Fe_8$ have, mostly, uniaxial magnetic anisotropy and the variation of the magnetic anisotropy barrier height $U_\pm$ on the external magnetic field applied parallel to the easy axis of magnetization is $U_\pm = U(1 \pm H/H_a)^2$.

The variation of the total magnetic moment, $M = M_+ - M_-$ on the external magnetic field is,

$$\frac{dM}{dH} = -\Gamma \frac{(M - M_{eq})}{r} \quad (2)$$

where $r = dH/dt$ is the sweeping rate of the external applied magnetic field, $\Gamma = \Gamma_+ + \Gamma_-$, $M_{eq}$ is the equilibrium magnetic moment which is given by

$$M_{eq}(H) = \frac{\Gamma_- - \Gamma_+}{\Gamma_- + \Gamma_+} M_S \quad (3)$$

and $M_S$ is the saturation magnetic moment of the sample.

The infinitesimal change in the temperature of a magnetic sample produced when a magnetic field is applied adiabatically is expressed as

$$dT = -\frac{T}{C_H} \left(\frac{\partial M}{\partial T}\right)_H dH \quad (4)$$

where $C_H$ is the heat capacity of the magnetic sample. The total magnetic entropy change $\Delta S_H$ of the magnetic system due to the application of the magnetic field is

$$\Delta S_H(T,H) = -\int_0^{H_{max}} \left(\frac{\partial M}{\partial T}\right)_H dH \quad (5)$$

where $H_{max}$ is the final applied magnetic field. This change in entropy is associated to the alignment of the spins in the system parallel to the magnetic field. Using equation (5) it is possible, therefore, to calculate exactly the entropy change in the molecular clusters as we know the theoretical expressions for both $M(T)$ and $M(H)$.

In the above expressions we have not considered the existence of quantum phenomena, that is, the relaxation due to the tunnel of the magnetic moment across the anisotropy barrier height[13]. The occurrence of the quantum tunneling of the magnetic moment in these molecules has been intensively discussed in the last years and for the purpose of this paper we are only interested in the modification we have to introduce in the variation of the barrier height $U$ with the field due to this quantum effect. The effective barrier height $U_{eff}(H)$ is the full barrier when the system is off resonance and a reduced barrier, due to tunneling, on resonance. This effective barrier, $U_{eff}(H,T)$, for both $Mn_{12}$ and $Fe_8$, has been experimentally determined[14-15],

$$U_{eff} = DS^2\left[\left|1-\frac{H}{H_a}\right|^2 - b\left|1-\left|\sin\boldsymbol{p}\frac{H}{H_n}\right|\right|^2\right] \qquad (6)$$

The values of $D$, $b$ and $H_a$ for both samples are given in Table I.

The samples we have studied consist of small crystallites of average length 10 μm ($Mn_{12}$) and 2 μm ($Fe_8$). The orientation of these samples was achieved by embedding the microcrystals in Araldit epoxy in a 5T magnetic field for 24 hours. The orientation was confirmed by performing magnetization measurements as a function of the magnetic field applied parallel and perpendicular to the orientation direction. Magnetic measurements have been carried out using a SQUID magnetometer.

In Figure 1 we show the low field, $H = 100$ Oe, magnetization for the two samples, $Mn_{12}$ and $Fe_8$, when they are first cooled in the absence of the applied field. The ZFC magnetization data are, therefore, due to the induced moment along the bias field causing the asymmetric change in the relaxation rates $\Gamma_+$ and $\Gamma_-$. The temperature at the maximum, $T_b= 3$K for $Mn_{12}$, see Figure 1b, and $T_b= 0.9$ K for $Fe_8$, see Figure 1a, corresponds to the blocking temperature above which the magnetic moment of the molecules behave superparamagnetically. In the inset of Figure 1b we show the variation with temperature of the termal remanent magnetization, TRM, for the $Mn_{12}$ sample which clearly goes to zero at $T=T_b$. The barrier height, $U = \ln(\boldsymbol{n}t)T_b \approx 25T_b$ (t, is the experimental resolution time, ν is the attempt frequency), of the magnetic anisotropy is, therefore, $U = 66$ K for $Mn_{12}$ and 27 K for $Fe_8$.

The calculation of the entropy change, $\Delta S_H$ has been derived from isothermal demagnetization measurements which have also been simulated using equations (2) and (3). In the case of magnetization measurements performed at small field and temperature intervals, $\Delta S_H$, can be approximated from equation (7)

$$|\Delta S_H| = \sum \frac{1}{T_{i+1}-T_i}\left(M_i - M_{i+1}\right)_H \Delta H_i \qquad (7)$$

Where $M_i$ and $M_{i+1}$ are the magnetization values measured in the field $H$ at temperature $T_i$ and $T_{i+1}$ respectively.

In Figure 2 we show the entropy variation for $Mn_{12}$ deduced for a field variation of 3T from saturation to zero field. The continuous line represents the theoretical values of the entropy variation deduced using (7) and simulated isothermal demagnetization curves. The agreement between they and experiment is remarkable good. We have also simulated, for both samples $Mn_{12}$ and $Fe_8$ clusters, the entropy variation for different values of the sweeping rate, $r$, of the applied magnetic field, $r = 0.01$ Hz, $r = 10$ Hz and $r = 100$ Hz, see Figure 3. From these results it is clearly deduced that the height of the entropy maximum depends on the sweeping rate of the applied field; in the case of $Mn_{12}$ the value 21 J/K·Kg is obtained for $r = 0.01$ Hz while at $r = 100$ Hz it is reduced to 13 J/K·Kg. The width of the entropy maximum decreases when decreasing the sweeping rate of the magnetic field and is of few K for both samples. All these experimental facts agree well with the occurrence of the magnetic blocking for all molecules at the same temperature. This blocking is, however, time dependent and consequently the entropy associated to this phenomenon should also change with time. In other words, the entropy variation should depend on the sweeping rate of the magnetic field in agreement with our experimental observation and calculations. Both the high value of the maximum variation of the entropy and the narrow distribution around this maximum suggest that these materials are good candidates to be used in magnetic refrigerators. Moreover, the possibility to shift the maximum of the entropy variation with the sweeping rate of the field, (or equivalently with the velocity of the sample entering the region where it is applied the field like in the case of the magnetic rotating device) suggest that we may use these materials to cover a reasonable broad interval of temperatures.

Finally we have estimated the magnetocooling effect as measured by the adiabatic temperature variation, $\Delta T$ associated to the entropy variation of Figures 2 and 3. This has been done by taking into consideration the temperature and field dependence of the heat capacity for both samples[16-17]. The large values obtained for $\Delta T$ indicate the possibility to cool down to the mK regime and to reduce the rate of evaporation of helium liquid.

In conclusion, these results indicate that the so called molecular clusters owing high spin and large magnetic anisotropy, have high potentiality to work as magnetic refrigerants in the helium liquid regime.

Acknowledgements.- This work was done in part due to the financial support of Carburos Metalicos S.A.

**Figure Captions.-**

**Figure 1.** Low field magnetization data for $Fe_8$, figure 1a, and $Mn_{12}$, figure 1b. The inset in figure 1b shows the temperature variation of the thermal remanent magnetization, TRM, for the $Mn_{12}$ sample.

**Figure 2.** Entropy variation for the $Mn_{12}$ sample deduced from isothermal demagnetization curves from $H = 3$ T to $H = 0$ T. The continuous line represents the theoretical fit of the experimental values. The value of the sweeping rate, $r = 7 \cdot 10^{-3}$ Hz, used in this fit corresponds to the time needed to switch off the magnetic field from 3 T.

**Figure 3.** Simulated entropy variation for different sweeping rates, $r = 0.01$ Hz (dotted line), $r = 10$ Hz (dashed line) and $r = 100$ Hz (continous line). The magnetic field change from 3 T to 0 T. (a) $Mn_{12}$ and (b) $Fe_8$.

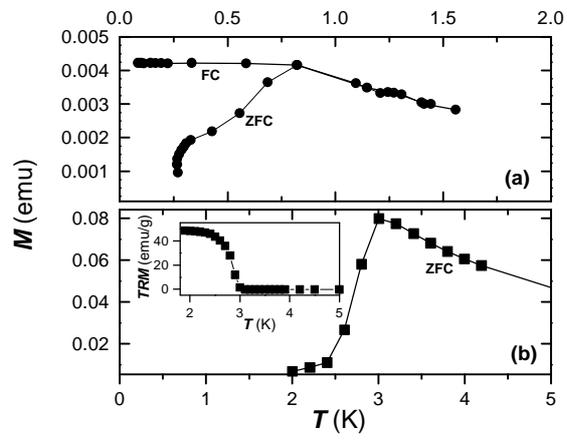

**Fig. 1, F. Torres, Appl. Phys. Lett.**

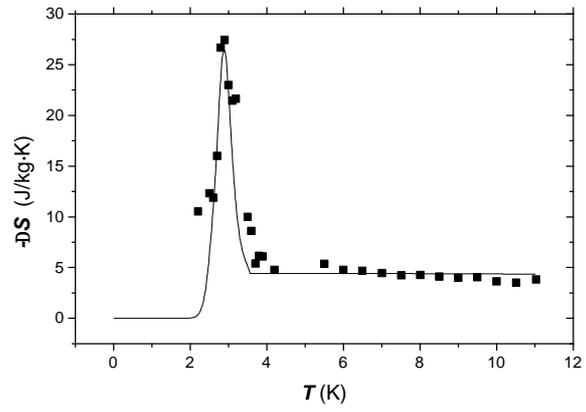

**Fig. 2, F. Torres, Appl. Phys. Lett.**

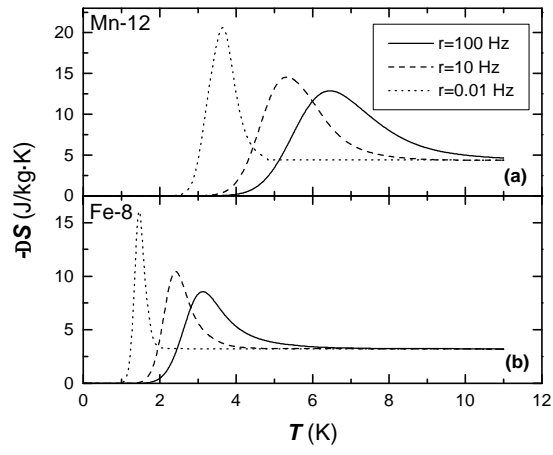

Fig. 3, F. Torres, Appl. Phys. Lett.